%% file: template.tex
\documentclass{IEEEcsmag}

\usepackage[colorlinks,urlcolor=blue,linkcolor=blue,citecolor=blue]{hyperref}

\usepackage{cite}
\usepackage{tabularx}
\usepackage{multirow}
\usepackage{makecell}
\usepackage{svg}

\usepackage{longtable}
\usepackage{booktabs}

\usepackage{upmath}
\usepackage{caption}
\usepackage{subcaption}
\usepackage{color}

\definecolor{Orange}{rgb}{0.8,0.4,0}
\definecolor{Blue} {rgb}{0,0,1}
\definecolor{Red} {rgb}{1,0,0}
\definecolor{Green} {rgb}{0,1,0}
\definecolor{Cyan} {rgb}{0,0.8,1}
\definecolor{Amethyst}{rgb}{0.6, 0.4, 0.8}

\jvol{XX}
\jnum{XX}
\paper{xx}
\jmonth{xxxx}
\jname{IEEE Special Issue on Special Issue on Human-Centered AI}
\pubyear{2022}

\begin{document}


\title{DDoD: Dual Denial of Decision Attacks on Human-AI Teams}

\author{Benjamin Tag}
\affil{The University of Melbourne}

\author{Niels van Berkel}
\affil{Aalborg University}

\author{Sunny Verma}
\affil{Macquarie University}

\author{Benjamin Zi Hao Zhao}
\affil{Macquarie University}

\author{Shlomo Berkovsky}
\affil{Macquarie University}

\author{Dali Kaafar}
\affil{Macquarie University}

\author{Vassilis Kostakos}
\affil{The University of Melbourne}

\author{Olga Ohrimenko}
\affil{The University of Melbourne}

\markboth{Department Head}{Paper title}

\begin{abstract}
Artificial Intelligence (AI) systems have been increasingly used to make decision-making processes faster, more accurate, and more efficient. However, such systems are also at constant risk of being attacked. While the majority of attacks targeting AI-based applications aim to manipulate classifiers or training data and alter the output of an AI model, recently proposed Sponge Attacks against AI models aim to impede the classifier's execution by consuming substantial resources. In this work, we propose \textit{Dual Denial of Decision (DDoD) attacks against collaborative Human-AI teams}. We discuss how such attacks aim to deplete \textit{both computational and human} resources, and significantly impair decision-making capabilities. We describe DDoD on human and computational resources and present potential risk scenarios in a series of exemplary domains.


\end{abstract}

\maketitle

\input{sections/1_introduction.tex}

\input{sections/2_background}
\input{sections/3_attacks}
\input{sections/4_discussion}

\input{sections/5_conclusion}
\input{sections/6_acknowledgements}

\bibliographystyle{IEEEtran}
\bibliography{references}

\input{sections/8_biographies}

\end{document}

%% file: sections/1_introduction.tex

\chapterinitial{Intelligent} machines have increasingly been integrated into human teamwork settings. Besides substituting human workers, Artificial Intelligence (AI) systems are deployed at a large scale to support humans\footnote{2020 Deloitte Global Human Capital Trends, \url{https://www2.deloitte.com/content/dam/Deloitte/cn/Documents/human-capital/deloitte-cn-hc-trend-2020-en-200519.pdf}, last accessed Sept. 27, 2022}. 
The efficiency and endurance of such machines effectively complement human capabilities. Therefore, rather than substituting workers with AI, an increasing number of companies are integrating it in collaborative Human-AI teams\footnotemark[\value{footnote}], to enable teams to work more effectively, improve decision-making, and increase productivity. By some accounts, the economic impact of this development is projected to add 13 Trillion USD to the global economy over the next ten years\footnote{\url{https://hbr.org/2019/07/building-the-ai-powered-organization}, last accessed March 29, 2022}. 

Unfortunately, such hybrid teams provide new attack vectors. While traditional attacks on humans often took the form of social engineering to steal credentials or gain secure access, we now have to consider attacks that target Human-AI teams aiming to deplete resources and impair their combined performance. 

In the machine learning (ML) domain, recent research has investigated the impact of the so-called Sponge Attacks on ML classifier performance~\cite{Shumailov2021SpongeExamples}. These attacks aim to occupy computational resources and obstruct the model's behavior and availability. In contrast to adversarial attacks that aim to distort predictions, sponge attacks aggravate decision-making by wasting resources while appearing disguised as rightful requests.

 In this article, we introduce \textit{Dual Denial of Decision (DDoD)} attacks by extending the notion of sponge attacks against human resources and advocate for an increased awareness of such attacks due to the risks they introduce to collaborative Human-AI teams. Such risks include confusing humans and presenting them with unclear choices, as well as flooding them with inconclusive classification outputs produced by the machine. This consequently results in a high demand for cognitive resources, increased cognitive load, and saturated attention, making humans more susceptible to other attacks, mistakes, and decreased performance~\cite{Sweller1994}. 

The Human-Computer Interaction (HCI) community has intensely worked to understand effective Human-AI collaborations better, focusing on explainability, transparency, and fairness. However, the robustness and efficacy of such Human-AI collaborations is yet to be fully explored. To unpack the potential impact of DDoD attacks on Human-AI teams, we analyze a series of traditional Human-AI collaboration scenarios and discuss the potential implications and risks of these attacks on their performance. To this end, we discuss implications for HCI, AI, Pervasive Computing, and the increasingly popular Human-AI research. We draw on the body of work in cybersecurity and Human-AI Interaction, paving the way for new research focusing on better protecting Human-AI teams.

%% file: sections/2_background.tex
\section{BACKGROUND}

We distinguish between three types of attacks that influence the performance of an ML model: \textit{Adversarial}, \textit{Poisoning}, and \textit{Backdoor} attacks.
The main difference between these attacks is the attacker's capability, i.e., where in the ML pipeline the attacker interferes: during training, inference (i.e., when a trained ML model is deployed to process data and produce predictions), or in both phases.

\textit{Adversarial attacks} harness benign samples that include small perturbations (e.g., noise on an image) to drastically alter a model's predictions~\cite{Goodfellow2015ExplainingAH}. An adversarial attack is executed at inference time without the need for involvement during training, instead exploiting inherent inconsistencies around the decision boundary from the training process. Such perturbations are typically imperceptible to humans, while they effectively deceive the model.

\textit{Poisoning attacks} manipulate a small proportion of the data used to train a model to significantly reduce the model's prediction performance on any input data~\cite{biggio2012poisoning}. Attackers can compromise a training dataset by submitting poisoned data at crowdsourcing, planting poisoned samples for data crawlers to collect, or contributing to training data directly (e.g., sending malware or spam emails to a system that collects any inputs it receives to retrain the models and stay abreast of evolving threats).

\textit{Backdoor attacks}, similar to poisoning attacks, have access to the training phase of the model, where an adversary can teach the model to behave in a pre-determined manner when a certain trigger is presented at run-time~\cite{chen2017targeted}. These triggers have evolved from static shapes to invisible noise, whilst also taking real physical forms in the world to fool image-based applications. 

All the above attack vectors have been studied extensively. However, typically these attacks are considered against a standalone ML model without humans, unlike collaborative Human-AI teams.

\subsection{Humans as the weak link}

When a human is part of the decision-making process, potential attacks on humans provide a new gateway for attackers to compromise collaborative Human-AI teams. Social engineering is a well-known technique where the human is attacked directly to obtain control of a system
(e.g., by enticing the user to follow fraudulent links in emails), leading to phishing, ransomware, or malware attacks. Social engineering describes the psychological manipulation of system users into unknowingly disclosing sensitive information and performing detrimental actions on behalf of the attacker, within the system~\cite{10.5555/1373319}. While machines can warn humans of such attacks (e.g., spam filters, system permission requests), the increasing deployment of Human-AI teams still presents underexplored avenues for attacks.

We note that while humans can leak private information (e.g., through phishing), complementary privacy attacks exist, whereby information about an ML model's training data is leaked~\cite{7958568}. However, in this article we focus on a new class of attacks against the integrity and robustness of the model, specifically attacks that change a model's behavior at inference or classification time, i.e., when the model is being deployed, and the impact such attacks can have in collaborative Human-AI teams.

\subsection{Sponge attacks}
Sponge attacks against ML models, introduced in~\cite{Shumailov2021SpongeExamples}, do not seek to compromise the predictive accuracy of the model like adversarial, poisoning, and backdoor attacks. Instead, a sponge attack provides inputs at inference time to drain the model's resources as much as possible, e.g., maximize the energy consumption and latency of inference by increasing the number of arithmetic operations or memory accesses required to process the input. 

\begin{figure*}[htb!]
  \centering
  \includegraphics[width=1.0\linewidth]{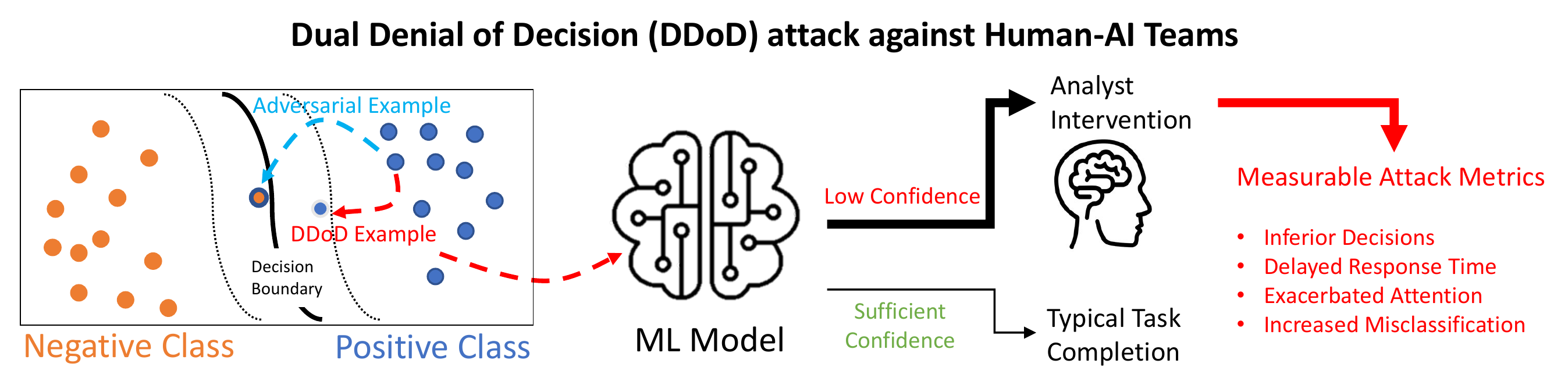}
  \caption{Schematic representation of a DDoD attack against Human-AI teams. In contrast to an adversarial attack (light blue) a DDoD example does not change the class of a data point, but rather moves it into a space of low classification confidence.}
  \label{fig:sponge}
\end{figure*}

The effectiveness of sponge attacks depends on the model, with Natural Language Processing (NLP) tasks shown to be particularly susceptible to such attacks. Shumailov~{et al.} demonstrate an attack on Microsoft Azure's translator, a real-world NLP system, to show an increase in response time from 1ms to 6s~\cite{Shumailov2021SpongeExamples}. Unlike denial of service attacks, sponge attacks increase the resource consumption of a system without increasing the number of requests to the system, thereby circumventing rate-limiting defenses. It is evident that sponge attacks have the potential for widespread negative effects on the availability and performance of ML models.

Along similar lines, Boucher~{et al.} consider sponge attacks in the context of text-based ML models and identify `imperceptible perturbations' as an attack vector that is highly challenging to spot for humans ~\cite{Boucher2022BadCharacters}. They demonstrate how text can be crafted such that it appears legitimate to humans, but deceives computers. The authors outline four means to achieve this: invisible characters ({e.g.}, zero-width spaces), homoglyphs (characters that look almost identical -- Cyrillic and Latin `A'), reorderings (use of control characters to alter the rendering order), and deletions (use of control characters to conceal characters within strings). Boucher~{et al.} present how a sponge attack can be deployed against search engines, machine translation systems, and NLP models, with potential consequences including the ability to bypass content detection systems and negatively impact training data. While their examples feature NLP applications, such sponge attacks can be extended to other applications like image processing or malware analysis. 

Sponge attacks have the potential to deceive machines whilst remaining hidden from humans. However, it remains to be verified whether such examples could be crafted not only to hide from the human but to concurrently influence \textit{both} the machine and the human. As such, the ability and effectiveness of Human-AI teams to sustain such attacks is a high-priority study area.

\subsection{Human-AI teams}

The way we design systems to enable successful collaboration in Human-AI teams has attracted increased interest in both the HCI and AI communities. Developing an understanding of end-user's expectations toward effective and efficient AI-powered collaborative agents is an active research challenge. For example, Zhang~{et al.} study the expectations of AI team members in a gaming context and describe that user expectations focus primarily on the AI's technical abilities, to provide the means to develop a shared understanding between the human user and the AI, and effective communication strategies~\cite{Zhang2021IdealHumanAI}. 

Integrating AI support in a real-world application requires a thorough understanding of, and adjustment to, the context in which it is deployed~\cite{Berkel2021Visual}. Examples of domains in which collaborative Human-AI systems have been studied include data science~\cite{Wang2019HAIDataScience}, clinical domains~\cite{Berkel2021Visual}, and creative tasks~\cite{Oh2018Cocreation}. Despite diverse application domains, a common thread across these studies is the conflict between the desire to comprehend the AI's suggestions and choices (often under the label of `explainability') while simultaneously avoiding undesired interruptions to the user while completing their tasks.

%% file: sections/4_discussion.tex
\section{Dual Denial of Decision Attacks on Human-AI Teams}

To better differentiate the characteristics of this attack from traditional sponge attacks and capture the impact on Human-AI teams, we introduce the term: \textit{Dual Denial of Decision \textbf{(DDoD)}} attacks. 

While sponge attacks were originally conceived to drain \emph{computational resources}~\cite{Shumailov2021SpongeExamples}, here we extend this notion to DDoD attacks, to be carried out against collaborative Human-AI teams to drain \emph{human resources}. As presented in Figure~\ref{fig:sponge}, a DDoD does not change the class of a data point, but rather moves it into a space of low classification confidence. Consequently, the AI calls in the human collaborator to make the final decision, which may be biased by the low confidence provided by the classifier. Hence, it is important to articulate the range and variety of Human-AI teams that can be affected. First, it is essential to consider who is the ultimate decision-maker in a Human-AI team: a \textit{human}, or a \textit{computer}? Second, it is relevant to identify the type of partnership within the team: is it \textit{monitoring} and \textit{collaboration}, or \textit{instruction}?

These two aspects help us identify a variety of Human-AI teams: 1) Human supervises computer, such as social media platforms moderation and filtering, airport security check, and passport control; 2) Computer supervises human, such as anti-virus software and driving safety systems; 3) Human controls computer, such as driving and remote-controlled UAVs; and 4) Computer controls human, such as warehouse employees picking up items, form-filling, and logins.

In this paper, we focus on asymmetric Human-AI teams, i.e., where the human is the ultimate decision-maker aided by AI decision-support. This is because such systems can easily scale up the computational resources, but not so much the human resources. Furthermore, depending on the type of Human-AI partnership, the attack can be either \textit{indirect} (during which the human steps in to resolve issues related to the AI) or \textit{direct} (whereby the attack aims specifically at deceiving and misleading the human). Due to their deceptive nature and ability to target systems as well as humans, we posit that DDoD attacks can substantially deteriorate the performance of Human-AI teams. 

Traditionally, social engineering attacks have long targeted humans as the weakest link in many computing systems, taking advantage of the scarcity of human cognitive resources. Short-term (i.e., working on timescales of minutes-hours) cognitive factors, such as vigilance (sometimes synonymously used with sustained attention), cognitive workload, and stress directly impact human susceptibility to fraud, deception, and distort their decision-making ability~\cite{Montanez2020}. We consider those as the primary DDoD attack vectors on Human-AI teams. 

\textit{Vigilance} describes the phenomenon of a fluctuating cognitive performance. This usually means that the longer a task demands cognitive effort, the more cognitive performance declines. Prior research has shown that performance  significantly declines over tasks lasting 30--60 minutes~\cite{Montanez2020}, making the human more prone to being less attentive to signs of fraud or deception. 

Closely related to vigilance and often interrelated, \textit{cognitive load} (CL) describes the cognitive demands tasks put on the performers. These demands mainly depend on the task complexity (intrinsic CL), the format of the information presented (extrinsic CL), and the person's processing effort (germane CL)~\cite{Pollock2002}. The sum of all three CL components describes the total CL of a human. Rather than diverting attention, an attacker could exhaust the human with undue CL by interfering with one of the three types or a combination thereof. 

Finally, changes in task load, declining attention, and unprecedented task demands (e.g., output from an attacked classifier) can lead to increased acute \textit{stress} in the human. While stress in short bursts can be beneficial~\cite{Montanez2020}, as it heightens attention, it often leads to an increased focus on the stress-inducing factor, and thus, to fewer attentional resources being available for other tasks and information.

We bring all these elements together by presenting a number of illustrative examples, reflecting a range of application areas to exemplify potential DDoD attacks on Human-AI teams.

\subsection{Scenarios}

In this subsection we present a set of examples of established Human-AI teams. Often mentioned in this regard is the \textit{control problem} which describes the failure of a human operator to detect malfunctioning machines due to complacency or over-reliance. Until now, it has been recommended to install collaborative Human-AI teams to counteract the control problem~\cite{Zerilli2019}. However, our examples of DDoD attacks show that Human-AI teams present new vulnerabilities. 

\paragraph{Medical Diagnosis}
AI methods have been increasingly applied in clinical environments~\cite{Berkel2021Visual}.
These are deployed for diagnostic, prognostic, and therapeutic purposes, primarily serving as a decision-support tool. That is, the AI does not have the authority to diagnose patients or determine treatment, but the output of the AI instead provides advice to a human clinician, who makes the decisions. This demonstrates an example of a collaborative Human-AI team, where an AI can assist the diagnosis process (e.g., by proposing specific tests for an accelerated determination of a medical condition or minimizing unnecessary testing). 

For example, as a decision-support tool for a human decision-maker a medical imaging AI can interpret images or videos and detect disease-specific symptoms. These may be highlighted in the images to streamline diagnostic decisions. A DDoD attack in this setting may entail increasing the uncertainty of the predicted diagnosis and highlighting wrong parts of the image. The latter will cause the clinician to waste precious time examining irrelevant parts of the image and potentially ordering unnecessary tests to reach a clear diagnosis. In the worst case, the increased uncertainty on the AI side may result in misleading the clinician in their diagnostic approach. Consequently, the clinician could go for the wrong tests losing crucial treatment time. Moreover, a lack of confidence expressed by a usually well-working AI, may lower the confidence and confuse the human. The clinician could call in for additional human support, which may be needed elsewhere.  

\paragraph{Law Enforcement}

Law enforcement often has to prioritize how to dedicate the limited human resources to enforce compliance. This has consequently led to the deployment of automated detection systems, supervised by humans, especially in the domain of traffic offenses~\cite{Rademacher2019}. Speeding, running red lights, and phone use while driving can be identified and captured by cameras. In most cases, an automatic plate recognition system would transcribe the plate number and issue an infringement to the registered vehicle owner. However, with varying environmental conditions, this recognition task may come with uncertainty requiring a human operator's intervention. 

In this scenario, a DDoD attack would entail perturbing the license plate to create uncertainty forcing a review by a human. For example, a sticker could be affixed on the car to cause the classifier to erroneously detect multiple license plates, or be uncertain about the car's actual license plate. In cases where the perturbation of the license plate image is to a level that the human is not able to clearly detect the correct alphanumerical combination, offenses may go unpunished. In large jurisdictions this may occupy a large number of human eyes, which will then be missing at other ends.

\paragraph{Passport Control - Immigration}

Another example is an immigration facility, e.g., at airports, where passengers are processed by an automated border control system, verifying the passport's authenticity, chip and biometric information. If any of these elements are in doubt, the individual is referred to a human border control officer. 

A DDoD attack on this system would seek to overwhelm the human officers by diverting bulk amounts of individuals away from passing through the automated system. These systems capture facial information with a camera, and in the process will also capture background information. Hence, one potential attack vector is the hijacking, or purchasing of electronic ad spaces to display adversarial/sponge examples to influence the biometric operation, or creating a backlog of passengers that require human management, thereby diverting security resources away from other sensitive areas. This may also delay processing, which can lead to increased stress among passengers and security personnel.

\paragraph{Semi-Automated Driving}

Semi-automated driving has emerged as a desirable feature and has been becoming increasingly prevalent in commercial vehicles. Its ability to perform well in both regular and atypical settings, whilst ensuring the safety of passengers and the public, are critically important features allowing to distinguish a market-leading system from competitors. If the AI senses uncertainty about navigating safely, the control is typically relinquished to the human driver. 

A DDoD attack would involve purposefully introducing uncertainty into the AI such that control frequently falls back to the human operator, even if not necessary. For instance, stickers or other visual decoys may be placed across a city, such that cars frequently hand-over control to the human driver, thereby degrading the perceived quality of such a system, and posing an additional burden to human drivers. This would not only impact comfort but potentially pose health risks to passengers as well as pedestrians, if the AI indicates a potential obstacle or problem that the driver has to immediately react to. In seemingly benign but high-speed situations this can have fatal consequences. Moreover, if the car frequently relinquishes control to the driver in such situations, drivers may lose trust in the system and refrain from using it.

\paragraph{Lending and Insurance}
Lending and Insurance companies have an inherent tolerance to risk. With AI, additional details about a customer's spending and behavioral habits can be automatically analyzed to illustrate reliable risk profiles~\cite{balasubramanian2018insurance}. An output may indicate the size of payments, or if the customer should be accepted at all. However, such outputs may be provided to a supervising human expert who makes the final decision. 

For example, consider an emergency loan program initiated to support businesses and individuals through an environmental disaster. Many such applications may be automatically processed to perform checks and balances to ensure only qualified parties are offered funding. If a human is integrated into the loop for the final decision, or to handle uncertain qualification status, this presents an opportunity for an attacker to waste the cognitive resources and time of these humans, thereby delaying the processing for legitimate applicants or rendering them illegitimate or questionable. For instance, certain keywords related to risk factors could be introduced into applications to increase the uncertainty of the classifier, which in turn flags more applications for human inspection. In an emergency situation such as mentioned above, this can lead to urgent cases not finding a human expert, thus, delaying potentially vital payments.

\paragraph{Recommender Systems}
Recommender systems are a type of human-facing AI deployed in many online scenarios~\cite{Zhang2021}, where human users may be faced with information overload. For example, in scenarios like online shopping, booking accommodations, selecting a movie to watch, or deciding on a restaurant to dine, users may find tens to thousands of appropriate options and, thus, struggle to find the optimal one. Recommenders are capable of helping the users by filtering out the least appropriate options and presenting a list containing a small number of best-matching items.

One could conceive a DDoD attack that, instead of listing the best options, presents a list of highly-similar items having a comparable probability to be selected by the user. In this attack, the compromised AI only increases the choice difficulty, as the burden of thoroughly examining the recommended options, identifying the subtle differences between them, and making the final selection is put on the user. This is likely to increase both the cognitive demand associated with the decision and the decision-making time.

\section{Implications and Considerations}
Whilst existing work on cybersecurity has primarily focused on technical flaws and attack surfaces, the introduction of Human-AI teams broadens the vector for attacks. Human vulnerabilities such as high cognitive load, high degree of stress, low degree of attentional vigilance, poor domain knowledge, or lack of experience, make us more susceptible to attacks~\cite{Montanez2020}.

\paragraph{Cognition}
In the literature, various definitions of fatigue, alertness, and performance can be found and differences of opinion exist about their meaning. For our purposes, the term performance comprises cognitive functions ranging in complexity from simple psycho-motor reaction time to logical reasoning, working memory, and complex executive functions. Fatigue refers to subjective reports of loss of desire or ability to continue performing. Alertness is a human resource that can be negatively influenced with carefully crafted sequences of inputs. One can see how an attack that leads to increased demand for human decision making can quickly deplete alertness. Since alertness is seen as a combination of selective attention, vigilance, and attentional control~\cite{VanDongen2000}, it describes the readiness to respond to stimuli and plays a role in higher cognitive functions affecting productivity, decision making
, and memory. 
Cognitive performance significantly declines for tasks lasting 30-60 minutes~\cite{Montanez2020}. Depleted levels of alertness, thus, lead to decreased productivity, slower or inhibited decision making, and an increased likelihood for mistakes. This can have particularly dangerous consequences, especially in situations where immediate reaction is required (e.g., traffic situations). 

\paragraph{Trust}
To safely integrate AI in organizations and society, it needs to be trusted by its users. A core component in building trust is the anticipation of intentional behavior in an entity, and the ability to predict (parts of) the decisions made by this entity and their impact~\cite{Jacovi2021}. Thus, trust guides the reliance on the output produced by increasingly complex AI that take on tasks that have a high cognitive demand on humans (e.g., predicting an illness within a split of a second based on thousands of images). This often happens without fully understanding the workings of the system. Consequently, when a system under a DDoD attack produces a large amount of results with low confidence, that can be either correct or wrong, the reliance on the system's performance deteriorates. The consequences of this are 1) trusting an unreliable system, which can result in false decisions, or 2) losing the advantages such a system has provided by not relying on it. 

To increase end-user trust in autonomous decision-making, AI research community has made extensive efforts to improve the \textit{explainability} of AI systems. While there is a great diversity in the intended use of such explanations (e.g., global vs local), explanations all face a similar challenge: they can interrupt a user's ongoing task and demand time and mental energy~\cite{Sweller1994, Berkel2021Visual}. As such, merely including explanations for the recommended actions is insufficient to support a Human-AI team in time-critical tasks such as averting DDoD attacks. Therefore, identifying the most relevant recommendations and actions is a critical component of making explainability meaningful in Human-AI collaboration.

\paragraph{Countermeasures}
Monitoring DDoD attacks on Human-AI teams, therefore, requires not only an assessment of system resource consumption, but also continuous and real-time monitoring of the human resource availability. The pervasive and ubiquitous computing community has provided a plethora of solutions in this domain. Through environmental and wearable sensors, cognitive load, attention, and stress can be detected. Prior work has shown that this data can be obtained unobtrusively~\cite{Fairclough2009}.

Once extensive human resource consumption has been detected, the system can proactively act to support the user by looking for ways to reduce their cognitive load. Such actions can range from suppressing interruptions to increasing the degree of AI support offered in the tasks presented.
In addition, a third source of information is the set of decisions made by the user. Here, historical information can reveal unexpected decision-making, which in turn can flag potential attacks. Similarly, decision-making can be used to assess the human's current abilities -- particularly in relation to the aforementioned detection of human resource consumption. Here we build on the concept of `golden questions' as frequently used in crowdsourcing scenarios -- a small set of tasks or questions to which the correct answer is known and can therefore be used to assess the quality of the worker's output. Such verification steps can be incorporated into Human-AI systems to monitor human performance.

\section{Future Research}
Despite the existing knowledge of monitoring human biosignals, a wide range of considerations remains open. While the user typically takes the role of the final decision-maker in many Human-AI systems, the proposed monitoring of the human's performance and decisions implies the user to be monitored by the AI. This raises challenging questions, particularly associated with interruption and communication of system state. Therefore, future research should investigate: 
\begin{itemize}
    \item The extent to which Human-AI teams are vulnerable to DDoD attacks.
    \item Parameters, signs, and indicators that enable Human-AI teams to detect DDoD attacks. 
    \item Potential defenses and ways to integrate DDoD defenses into the design of Human-AI systems from the beginning and not as an afterthought. 
    \item How to make the human collaborator cognitively less vulnerable to overload.
    \item How the output of a system under attack influences factors such as trust in the system. 
    \item How decisions made by a human collaborator under attack influence the ML model. 
    \item How to mitigate the control problem leading to human collaborators blindly trusting the AI output. 
    \item Format, design, and timing of explainability elements that help lowering the cognitive demand on the human.
\end{itemize}

%% file: sections/5_conclusion.tex
\section{CONCLUSION}
The HCI community's extensive efforts toward establishing a better understanding of effective Human-AI collaborations are far from complete. Among colossal challenges such as explainability, transparency, and fairness, security and robustness have largely remained an underexposed element of collaborative Human-AI systems. Preparing Human-AI teams for sponge and DDoD attacks requires a thorough understanding of both technical and human limitations, as well as cross-disciplinary collaboration. Building on established HCI knowledge, future research could explore how to design and implement interactive systems that monitor both AI and human performance -- alerting and supporting when either party faces atypical resource consumption. This article highlights how DDoD attacks can have severe consequences on the performance of collaborative Human-AI teams and calls for active cross-disciplinary research on this emerging topic.

%% file: sections/6_acknowledgements.tex
\section{ACKNOWLEDGMENT}

This work was supported by the joint CATCH MURI-AUSMURI.

%% file: sections/8_biographies.tex
\section{BIOGRAPHIES}
\label{bios}

\begin{IEEEbiography}{Benjamin Tag}{\,}is a Postdoctoral Research Fellow at the School of Computing and Information Systems at the University of Melbourne. He researches digital Human-AI Interaction, emotion regulation, and human cognition, with a special focus on inferring mental state changes from biophysical signals in the wild. Contact him at benjamin.tag@unimelb.edu.au.
\end{IEEEbiography}

\begin{IEEEbiography}{Niels van Berkel}{\,}is an Associate Professor at Aalborg University in Denmark. His research interests focus on Human-Computer Interaction, Social Computing, and Human-AI interaction. Contact him at nielsvanberkel@cs.aau.dk.
\end{IEEEbiography}

\begin{IEEEbiography}{Sunny Verma}{\,} is a Postdoctoral Research Fellow at Macquarie University. Prior to that, he has worked as a postdoctoral fellow at the Data Science Institute, UTS. His research interests are Fairness and Human Cognition in AI systems and data mining. Contact him at sunny.verma@mq.edu.au.
\end{IEEEbiography}

\begin{IEEEbiography}{Benjamin Zi Hao Zhao}{\,} is a Postdoctoral Research Fellow at Macquarie University. His current research interests are authentication systems, security and privacy attacks against machine learning, and rapid malware triage.
Contact him at ben\_zi.zhao@mq.edu.au.
\end{IEEEbiography}

\begin{IEEEbiography}{Shlomo Berkovsky}{\,} is a Professor at Macquarie University. He leads the Precision Health stream at the Centre for Health Informatics. The stream focuses on the use of AI methods to develop patient models and personalized predictions of diagnosis and care, and on the ways clinicians and patients interact with health technologies.
Contact him at shlomo.berkovsky@mq.edu.au.
\end{IEEEbiography}


\begin{IEEEbiography}{Dali Kaafar}{\,} is a Professor of cyber security and privacy-preserving technologies at the Faculty of Science and Engineering. He is also the executive director of the Macquarie University Cyber Security Hub. His research interests include digital privacy, distributed systems security, authentication systems and security risks measurement and modeling. Contact him at dali.kaafar@mq.edu.au.
\end{IEEEbiography}

\begin{IEEEbiography}{Vassilis Kostakos}{\,}is a Professor of Computer Science at the University of Melbourne in Australia where he leads the Human-Computer Interaction Group. His research interests focus on ubiquitous computing, human-computer interaction, social computing, and the Internet of Things. Contact him at vassilis.kostakos@unimelb.edu.au.
\end{IEEEbiography}

\begin{IEEEbiography}{Olga Ohrimenko}{\,}is an Associate Professor at the University of Melbourne.
Her research interests include privacy and security of machine learning and data analysis.
Contact her at oohrimenko@unimelb.edu.au.
\end{IEEEbiography}